\documentclass[conference]{IEEEtran}
\IEEEoverridecommandlockouts
\usepackage{array}
\newcolumntype{P}[1]{>{\centering\arraybackslash}p{#1}}
\usepackage{multirow}
\usepackage{cite}
\usepackage{makecell}

\usepackage{amsmath,amssymb,amsfonts}
\usepackage{algorithmic}
\usepackage{graphicx}
\usepackage{textcomp}
\usepackage[dvipsnames]{xcolor}
\usepackage{xcolor}
\usepackage[hidelinks]{hyperref}

\def\BibTeX{{\rm B\kern-.05em{\sc i\kern-.025em b}\kern-.08em
    T\kern-.1667em\lower.7ex\hbox{E}\kern-.125emX}}
\begin{document}

\title{Exposing LLM Vulnerabilities: Adversarial Scam Detection and Performance }

\author{\IEEEauthorblockN{Chen-Wei Chang\IEEEauthorrefmark{1},
Shailik Sarkar\IEEEauthorrefmark{1},
Shutonu Mitra\IEEEauthorrefmark{1},
Qi Zhang\IEEEauthorrefmark{1}, 
Hossein Salemi\IEEEauthorrefmark{2},
Hemant Purohit\IEEEauthorrefmark{2},
Fengxiu Zhang\IEEEauthorrefmark{3}, \\
Michin Hong\IEEEauthorrefmark{4},
Jin-Hee Cho\IEEEauthorrefmark{1},
Chang-Tien Lu\IEEEauthorrefmark{1},
}
\IEEEauthorblockA{\IEEEauthorrefmark{1} Department of Computer Science, Virginia Tech, USA}
\IEEEauthorblockA{\IEEEauthorrefmark{2}Department of Information Sciences and Technology,\IEEEauthorrefmark{3}School of Policy and Government, George Mason University, USA}
\IEEEauthorblockA{\IEEEauthorrefmark{4}School of Social Work, Indiana University, USA}
}
\maketitle

\begin{abstract}
Can we trust Large Language Models (LLMs) to accurately predict 
scam? 
This paper investigates the vulnerabilities of LLMs when facing adversarial scam messages for the task of scam detection. 
We addressed this issue by creating a comprehensive dataset with fine-grained labels of scam messages, including both original and adversarial scam messages. 
The dataset extended traditional binary classes 
for the scam detection task into more nuanced scam types. Our analysis showed how adversarial examples took advantage of vulnerabilities of a LLM, 
leading to high misclassification rate. We evaluated the performance of LLMs on these adversarial scam messages and proposed strategies to improve their robustness.
\end{abstract}

\begin{figure*}[t]
    \centering
    \includegraphics[width=\textwidth]{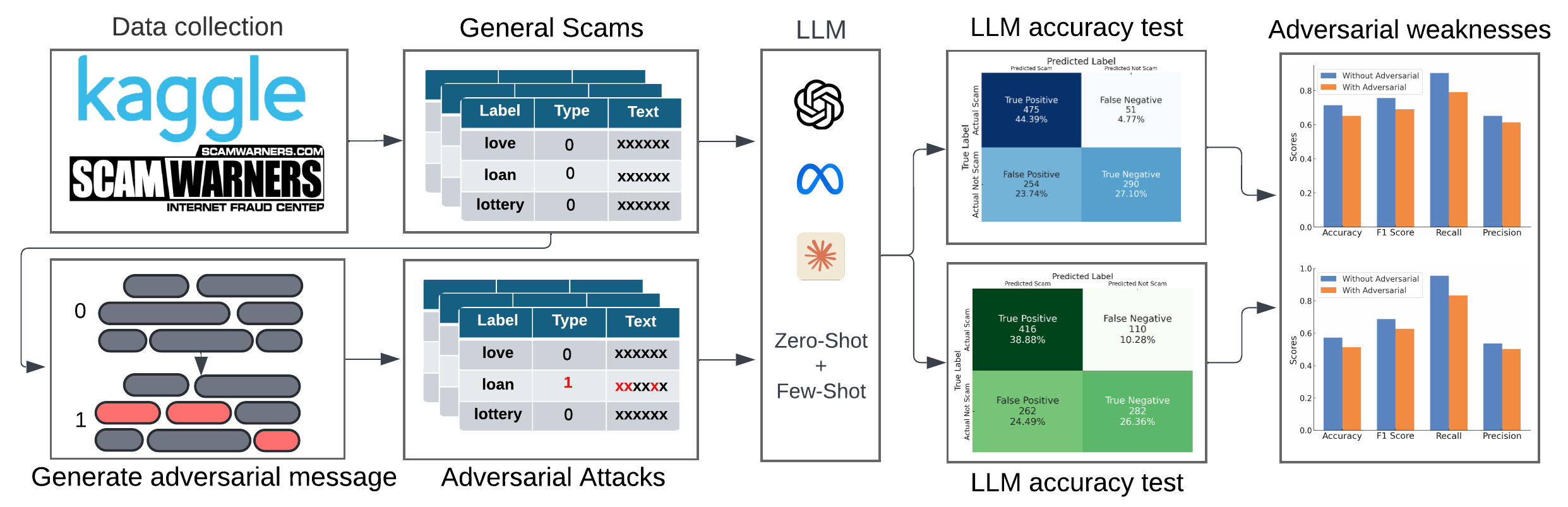} 
    \vspace{-7mm}
    \caption{Experimental procedures, including data collection, annotation, LLM testing, and result analysis.}
    \label{fig1}
    \vspace{-3mm}
\end{figure*}

\section{Introduction}
Scams are becoming increasingly sophisticated, targeting vulnerable groups such as job seekers, the elderly, and individuals seeking relationships. Detecting scams is crucial to prevent financial loss and data breaches. Large Language Models (LLMs) have gained popularity for scam detection~\cite{shen2024combating} due to their ability to comprehend complex language of text-based 
messages. However, these models remain susceptible to adversarial examples, where small alterations in the text can deceive the model, leading to incorrect classifications. While current LLM-based systems perform well with straightforward scams, they struggle with adversarially or artificially crafted scam messages~\cite{salman2022empirical}, resulting in higher misclassification rates and exposing individuals to significant risks. 

This paper aims to explore the vulnerabilities of Large Language Models (LLMs) in detecting adversarial scam messages, where original scam messages are strategically modified based on patterns recognized by the LLMs. By focusing on how adversarial examples leads to exploitation of the weaknesses or vulnerabilities of LLMs, this research aims to shed light on the limitations of current models. The {\bf key contributions} of this work are as follows:
\begin{itemize}
\item \textbf{Curating a comprehensive scam dataset with fine-grained labels and adversarial examples:} We developed a nuanced labeling scheme and generated adversarial examples designed to deceive LLMs, creating a dataset for investigating model vulnerabilities and evaluating LLM performance under adversarial conditions. 
    
\item \textbf{Identifying LLM vulnerabilities to adversarial examples in scam detection:} Our framework uses the labeled dataset to assess model robustness in low-shot learning (zero-shot and few-shot~\cite{hu2022can})-based adversarial settings. We evaluated LLMs by comparing accuracy on original versus adversarial scam messages, revealing the extent of misclassification and highlighting the need for stronger detection methods.
    
\item \textbf{Evaluating LLM performance on scam messages to assess vulnerabilities:} We further explored strategies to counter adversarial data augmentation
, and our results indicate that specific adversarial prompting techniques can help mitigate such attacks.
\end{itemize}

\section{Related work}


{\bf Traditional scam detection algorithms.} Decision trees~\cite{dileep2021novel} and support vector machines (SVMs)~\cite{modupe2011exploring} have been effective in phishing and scam message detection, but they still face significant limitations. The advent of LLMs has transformed text classification~\cite{sun2023text}, providing a more nuanced understanding and enhanced capability for detecting threats~\cite{jiang2024detecting}. However, traditional algorithms often struggle with adapting to adversarial inputs and handling the complexity of evolving scam tactics, limiting their effectiveness in real-world scenarios.

\vspace{1mm}
{\bf Vulnerabilities of LLMs in scam detection.} Despite their success in many NLP tasks, LLMs remain vulnerable to adversarial examples in scam detection~\cite{lin2024large}. Small input modifications can easily mislead these models into incorrect classifications. Even minor changes can deceive advanced models like GPT-3.5, leading to inaccurate outputs~\cite{xu2023llm, khatun2024uncovering}. These vulnerabilities underscore the need for stronger mechanisms to fortify LLMs against adversarial attacks in practical applications.

The vulnerability of LLMs to adversarial attacks has been widely recognized~\cite{kulkarni2024ml}, with research showing that adversarial text can effectively deceive these models~\cite{raina2024llm}. This raises concerns about the reliability of current scam detection systems in security-critical environments~\cite{salman2022empirical}. 

To fill the gaps, this work examines how adversarial scam messages exploit these weaknesses. We also emphasize the need for more resilient detection methods~\cite{mo2013resilient} to ensure reliable performance in real-world applications.

\section{Experimental Setup \& Methodology}

\subsection{Dataset Generation}
As shown in Table~\ref{tab:dataset-example}, we compiled a dataset of approximately 1,200 messages, manually labeled from sources, such as Kaggle~\cite{fraudulent_job_posting_2023, email_spam_2023, spam_or_not_spam_2023} and Scamwarners~\cite{scamwarners_website}, representing various types of scam and non-scam content. The dataset is categorized into three main groups: original scam messages (530), adversarially modified scam messages (126), and non-scam messages (544). The example dataset in Table~\ref{tab:dataset-example} illustrates a recruitment message. The first row presents an original recruitment scam message (i.e., an unmodified scam), followed by an adversarial recruitment scam message (i.e., a scam message altered based on LLM-detected patterns), and finally, a non-scam recruitment message.

\begin{table}[h]
\centering
\caption{\small Examples of Dataset Generation} \label{tab:dataset-example}    
\vspace{-2mm}
\scalebox{0.8}{
\begin{tabular}{|P{2cm}|p{8cm}|} \hline 
         Label  & Text\\ \hline 
         Original recruitment scam message & This is to inform you that we are currently hiring foreign, international, reputable, and experienced applicants for various job positions available. If you are interested, kindly apply by sending your CV/résumé to james.chevron.hr@job4u.com for more details.

Regards
James Armitage
EMEA Recruitment Team Lead.
HR Chevron Corporation London UK\\ \hline 
         Adversarial recruitment scam message  & This is to inform you that we are currently hiring international students and experienced applicants for the  \textcolor{RawSienna}{Software Engineer intern positions.     During our 10-week, paid summer internship program, you can contribute to meaningful work with supportive colleagues. You'll also benefit from: One-on-one mentorship.  Networking and social activities.  Training and workshops.  Exclusive speaker series  Professional development opportunities}    If you are interested, kindly apply by  \textcolor{RawSienna}{replying to this email} with your CV/résumé for more details    Regards,  James Armitage    EMEA Recruitment Team Lead\\ \hline 
         Non-scam recruitment message  & 
Hey Ashley,
Can you please confirm your availability for the meeting next week? We need to finalize the schedule. 

Cheers,
Tracie Gutierrez\\ \hline
    \end{tabular}}
\end{table}

To assess the robustness of LLMs in scam detection, we designed a series of adversarial scam messages using a structured methodology. These adversarial examples were generated by providing specific prompts to GPT-4, following the principles outlined below:

\begin{itemize}
    \item \textbf{Element Removal or Modification}: We removed elements commonly identified as scam indicators, such as payment instructions or unusual information requests.
    \item \textbf{Tone and Language Adjustment}: The tone was made more formal or neutral, reducing emotional language.( e.g., “urgent,” “exclusive,” and “limited-time offer,”)

    \item \textbf{Preservation of Key Information}: Key details (e.g., rental, shopping, or investment terms) were retained but reordered to appear less direct.
    \begin{itemize}
        \item Original: ``We fund viable projects at 2\% interest rate for 1-10 years.''
        \item Modification: ``Competitive rates of 2-6\% for loan terms ranging from 1-10 years.''
    \end{itemize}
    \item \textbf{Prompt Example}:You are an assistant helping to generate subtle versions of scam messages for research purposes. 
Follow these guidelines for all rewrites:
1. Format: Generate only the message content without any subject line or email header
2. Remove Obvious Scam Indicators: No urgent requests or unusual payment demands
3. Adjust Tone: Use professional, neutral tone
4. Retain Key Content: Keep core information but phrase legitimately
5. Add Limited Credibility: Include general references to known locations/institutions sparingly
\end{itemize}
The resulting adversarial scam messages preserved the original semantic content while reducing the likelihood of being classified as scams by LLMs.

\subsection{Experimental Methodology}
To evaluate the effectiveness of different models and prompt settings, we used two datasets: \textbf{General Scams} (unmodified scam messages) and \textbf{Adversarial Scams} (modified scam messages). We conducted experiments with three LLMs, including GPT-3.5, {\em Claude3-haiku}, and {\em LLaMA 3.1 8B Instruct}, and compared their scam detection performance under various datasets and learning techniques.

\begin{table*}[t]
    \centering

\caption{Performance Comparison of LLMs Across Different Scam Categories Using Various Prompt Types and Datasets}
\label{tab:scam-categories}
\renewcommand{\arraystretch}{1.3}
\resizebox{1\textwidth}{!}{
\begin{tabular}{|c|c|c|c|c|c|c|c|c|}
        \hline
        \/\multirow{2}{*}{Model} & \multirow{2}{*}{Prompt Type} & \multirow{2}{*}{Dataset}  &   \multicolumn{2}{c|}{Romance} & \multicolumn{2}{c|}{Finance} & \multicolumn{2}{c|}{Recruitment} \\
        \cline{4-9}
        \ &  &  & Accuracy & F1 Score & Accuracy & F1 Score & Accuracy & F1 Score \\
        \hline
        \multirow{3}{*}{GPT 3.5 Turbo} & \multirow{2}{*}{Few-shot using regular scam} & General Scams& 0.92& 0.91& 0.88& 0.86& 0.79& 0.71\\
        \cline{3-9}
        &  & \multirow{2}{*}{Adversarial Attacks}& 0.81& 0.75& 0.79& 0.71& 0.75& 0.64\\
        \cline{2-2} \cline{4-9}
        & Few-shot using adversarial scam &  & 0.81& 0.75& 0.87& 0.86& 0.77& 0.72\\
        \hline
        \multirow{3}{*}{Claude3-haiku} & \multirow{2}{*}{Few-shot using regular scam} & General Scams& 0.81& 0.81& 0.83& 0.84& 0.72& 0.71\\
        \cline{3-9}
        &  & \multirow{2}{*}{Adversarial Attacks}& 0.70& 0.66& 0.77& 0.77& 0.66& 0.62\\
        \cline{2-2} \cline{4-9}
        & Few-shot using adversarial scam &  & 0.62& 0.60& 0.77& 0.80& 0.67& 0.66\\
        \hline
        \multirow{3}{*}{LLaMA 3.1 8B Instruct} & \multirow{2}{*}{Few-shot using regular scam} & General Scams& 0.83& 0.8& 0.92& 0.91& 0.80& 0.71\\
        \cline{3-9}
        &  & \multirow{2}{*}{Adversarial Attacks}& 0.73& 0.64& 0.81& 0.76& 0.74& 0.60\\
        \cline{2-2} \cline{4-9}
        & Few-shot using adversarial scam &  & 0.62& 0.57& 0.68& 0.66& 0.63& 0.59\\
        \hline
    \end{tabular}
    }
\end{table*}

\section{Experimental Results \& Analyses}

\subsection{Performance Comparison of LLMs on General and Adversarial Scam Detection Across Different Categories}

\autoref{tab:scam-categories} presents the performance comparison of three LLMs, which are {\em GPT-3.5 Turbo}, {\em Claude3-haiku}, and {\em LLaMA 3.1 8B Instruct}, across different scam categories (Romance, Finance, and Recruitment) under various prompt types and datasets (general scams and adversarial scams). The results indicate that {\em GPT-3.5 Turbo} consistently outperforms the other models across all categories, demonstrating higher resilience to adversarial modifications. In contrast, {\em LLaMA 3.1 8B Instruct} shows the weakest performance, particularly when exposed to adversarial scam messages, likely due to its smaller parameter size and limited capacity to learn complex patterns. 

The \textbf{difference between categories} reveals that Romance scams are the most vulnerable to adversarial modifications, likely because the adversarial prompts are finance-related, leading models to misclassify emotional manipulations characteristic of Romance scams. While the few-shot setting with adversarial prompts helped partially recover performance across all models, the recovery was more pronounced for {\em GPT-3.5 Turbo}, indicating its superior ability to adapt to adversarial cues. Overall, the results emphasize that additional methods, such as adversarial training, are needed to enhance robustness, particularly for smaller models like LLaMA 3.1 8B.

\subsection{Performance Comparison of LLMs on General and Adversarial Scam Detection}

\autoref{tab:combined_performance} presents the performance comparison of LLMs, including {\em GPT-3.5 Turbo}, {\em Claude3-haiku}, and {\em LLaMA 3.1 8B Instruct}, evaluated on two datasets: general scams and adversarial scams. The models were tested using both zero-shot and few-shot prompt settings, with performance metrics including accuracy, precision, recall, and F1 score.

The results show that {\em GPT-3.5 Turbo} consistently outperforms the other models across both datasets. Its performance improves significantly in the few-shot setting, where regular scam examples boost detection capabilities. For general scams, its accuracy rises from 0.71 (zero-shot) to 0.87 (few-shot), and for adversarial attacks, from 0.79 to 0.83. {\em GPT-3.5 Turbo} also achieves higher F1 scores, reflecting its ability to handle both straightforward and adversarially modified scam messages.

Claude3-haiku also shows moderate performance improvement in the few-shot setting, although it lags behind {\em GPT-3.5 Turbo}. For general scams, its accuracy increases from 0.53 (zero-shot) to 0.69 (few-shot). However, Claude3-haiku experiences a notable drop in performance when confronted with adversarial examples, indicating its weaker resilience to such modifications. 

{\em LLaMA 3.1 8B Instruct} shows the weakest performance overall, especially when exposed to adversarial scams. While its accuracy improves slightly with few-shot prompts (from 0.57 to 0.69 for general scams), it struggles with adversarial examples, where accuracy drops to 0.59. This suggests that LLaMA’s smaller parameter size limits its ability to handle complex adversarial features compared to other models.

Overall, the findings emphasize that while few-shot prompting can improve performance in non-adversarial settings, additional methods such as adversarial training~\cite{shafahi2019adversarial} are needed to enhance robustness, particularly for smaller models like LLaMA 3.1 8B.

\begin{table*}[t]
\centering
\renewcommand{\arraystretch}{1.2}
\caption{Performance Analysis of LLMs in Scam Detection Under Various Prompt Types and Dataset Conditions}
\label{tab:combined_performance}
\begin{tabular}{|c|c|c|c|c|c|c|}
    \hline
    {\textbf{Model}}& {\textbf{Prompt Type}} &{\textbf{Dataset}} & \textbf{Accuracy} & \textbf{Precision} & \textbf{Recall} & \textbf{F1 Score}  \\ \hline
    \multirow{4}{*}{GPT-3.5 Turbo}& Zero-shot                           & \multirow{2}{*}{General Scams}& 0.71              & 0.65               & 0.90             & 0.76              \\ \cline{2-2} \cline{4-7}
                                      & \multirow{2}{*}{Few-shot using regular scam}&                                    & 0.87              & 0.90               & 0.82             & 0.86              \\ \cline{3-7}
                                      &                                    & \multirow{2}{*}{Adversarial Attacks}& 0.79              & 0.88               & 0.67             & 0.76              \\ \cline{2-2} \cline{4-7}
                                      & Few-shot using adversarial scam&                                    & 0.83              & 0.88               & 0.77             & 0.82              \\ \hline
    \multirow{4}{*}{Claude3-haiku}    
                                      & Zero-shot                           & \multirow{2}{*}{General Scams}& 0.53              & 0.51               & 0.97             & 0.67              \\ \cline{2-2} \cline{4-7}
                                      & \multirow{2}{*}{Few-shot using regular scam}&                                    & 0.69              & 0.62               & 0.85             & 0.72              \\ \cline{3-7}
                                      &                                    & \multirow{2}{*}{Adversarial Attacks}& 0.68              & 0.66               & 0.72             & 0.69              \\ \cline{2-2} \cline{4-7}
                                      & Few-shot using adversarial scam&                                    & 0.70              & 0.66               & 0.82             & 0.73              \\ \hline
    \multirow{4}{*}{LLaMA 3.1 8B Instruct} 
                                      & Zero-shot                           & \multirow{2}{*}{Non-Adversarial}   & 0.57              & 0.53               & 0.96             & 0.68              \\ \cline{2-2} \cline{4-7}
                                      & \multirow{2}{*}{Few-shot using regular scam}&                                    & 0.69              & 0.62               & 0.85             & 0.72              \\ \cline{3-7}
                                      &                                    & \multirow{2}{*}{Adversarial}       & 0.62              & 0.58               & 0.70             & 0.63              \\ \cline{2-2} \cline{4-7}
                                      & Few-shot using adversarial scam&                                    & 0.59              & 0.51               & 0.73             & 0.60              \\ \hline
\end{tabular}
\vspace{-8pt}
\end{table*}

\subsection{Case Study}
In this case study, we examine how the original scam message was modified to create an adversarial version that bypasses detection. The \textbf{original scam message} is as follows:

\begin{quote}
“We are an investment and loan financing group. We fund economically viable projects at 2\% interest rate for 1-10 years and 6-12 months grace period. Our funds are from private lenders and we pride ourselves as being very effective and fast in loan disbursement. I can be reached on email and Whatsapp: +97155 647 4204. Contact us for more details. Regards, MA, Financial Consultant.”
\end{quote}

\textbf{Adversarial Version:}

\begin{quote}
\textcolor{RawSienna}{Hello}, We are an investment and loan financing group \textcolor{RawSienna}{located in New York City.} We fund economically viable projects with interest rates ranging from \textcolor{RawSienna}{2\% to 8\%} for terms of 1-10 years and offer a grace period of 6-12 months, \textcolor{RawSienna}{varying by situation}. Our funds are sourced from private lenders, and we pride ourselves on our effectiveness, speed, and transparency in loan disbursement. For more details, please contact me \textcolor{RawSienna}{via email at sseanh@gmail.com} or on WhatsApp at +571 258 3824. \textcolor{RawSienna}{Thank you. Kind regards}, James Armitage, MA, Financial Consultant.
\end{quote}
In this adversarial version, several key changes were made:

\begin{itemize}
    \item \textbf{Addition of location information}: The phrase “located in New York City” was added to enhance credibility, making the message appear less suspicious.
    \item \textbf{Modification of interest rates and grace period}: The interest rate was expanded to a range (“2\% to 8\%”), and the grace period was made to seem more flexible.
    \item \textbf{Language adjustment}: The tone was made more formal and professional, reducing emotional cues or pressure typically present in scam messages.
    \item \textbf{Contact information update}: The original contact number and email were replaced with more generic options to bypass detection patterns linked to the original scam.
\end{itemize}

The LLM misclassified the modified message as non-scam due to the addition of location details, the expanded interest rate range, and the removal of high-risk terms. This highlights how adversarial examples can exploit an LLM’s reliance on specific keywords, emphasizing the need for future models to incorporate deeper contextual understanding and more sophisticated pattern recognition.

\section{Conclusion}
This work explores vulnerabilities in Large Language Models (LLMs) for scam detection by analyzing their performance on adversarial scam messages. Our results showed that even small modifications significantly reduced LLM accuracy. Experiments with models like GPT-3.5, Claude 3, and LLaMA 3.1 demonstrated decreased performance against adversarial examples. We also developed a dataset of original and adversarial scam messages across various scam types. To improve LLM robustness, we proposed strategies like adding adversarial prompts and using few-shot learning. Our findings emphasize the need to continually enhance LLM training methods to build a more resilient scam detection system.

\section*{Acknowledgement}
This work is partly supported by the Commonwealth Cyber Initiative (CCI) through its Inclusion and Accessibility in Cybersecurity program.
\bibliographystyle{IEEEtran}
\bibliography{ref}

\vspace{12pt}

\end{document}